\documentclass[prd,preprint,nofootinbib]{revtex4}
\usepackage{graphicx}
\usepackage{amssymb}
\usepackage{amsmath}

\begin{document}

\title{Power Spectrum of the Density Perturbations From
Smooth Hybrid New Inflation Model}
\author{Masahiro Kawasaki}
\author{Tsutomu Takayama}
\affiliation{Institute for Cosmic Ray Research,
The University of Tokyo,
Kashiwa 277-8582, Japan
}
\author{Masahide Yamaguchi}
\affiliation{Department of Physics and Mathematics, 
Aoyama Gakuin University, 
Kanagawa 229-8558, Japan}
\author{Jun'ichi Yokoyama}
\affiliation{Research Center for the Early Universe (RESCUE), 
Graduate School of Science, The University of Tokyo, 
Tokyo, 113-0033, Japan}
\date{\today}

\begin{abstract}
We numerically investigate density perturbations 
generated in the smooth hybrid new inflation model,
a kind of double inflation model that is designed to reproduce 
the running spectral index suggested by the WMAP results.
We confirm that this model provides the running spectral index
within 1$\sigma$ range of the three year WMAP result.
In addition, we find a sharp and strong peak on the spectrum of 
primordial curvature perturbation at small scales.
This originates from amplification of fluctuation in the first inflaton fields 
due to parametric resonance, 
which takes place in the oscillatory phase between two inflationary regime.
Formation probability of primordial black holes (PBHs) is discussed 
as a consequence of such peak.
\end{abstract}

\pacs{98.80.Cq}
\maketitle

%

%%%%%%%%%%%%%%%%%%%%%%%%%%%%%%%%%%%%%%%%
\section{Introduction}
%%%%%%%%%%%%%%%%%%%%%%%%%%%%%%%%%%%%%%%%%%%

The observation of Wilkinson Microwave Anisotropy Probe (WMAP) has
successfully determined the cosmological parameters and nature of 
density fluctuations with high precision. The WMAP result is 
quite consistent with the prediction of the inflationary cosmology, {\em{i.e.}},
the flat universe with almost scale-invariant adiabatic density fluctuation.
If one take a closer look at the shape of the power spectrum of
the density fluctuations obtained by WMAP, however, small deviation from 
scale invariance is found. In fact, the WMAP 1st year result suggested
that the data fits to the power spectrum with running spectral 
index $n_s$~\cite{Spergel:2003cb,Peiris:2003ff}, although the
statistical significance was not high enough 
\cite{Slosar:2004xj}. There has been a renewed interest in the running spectral
index because it is favored by the recently published three year WMAP 
result~\cite{Spergel:2006hy} which gives 
$dn_s/d\ln k = -0.055^{+0.029}_{-0.035}$. 
In Ref.~\cite{Seljak:2004xh},
no evidence of running was found from the combining analysis 
of the 1st year WMAP and  the improved data of Ly-$\alpha$ forests. 
However, as mentioned in~~\cite{Spergel:2006hy},  the three year WMAP result 
is not in good agreement with this
Ly-$\alpha$ data. Then, it seems  premature to adopt the 
power spectrum obtained from Ly-$\alpha$ forests. Therefore
it is worth studying the inflation model which provides the running 
spectral index.  
 
It is not an easy task to build an inflation model which produces
density perturbations whose power spectrum has a running index
\cite{Ballesteros:2005eg,Chen:2004nx}. 
After the release of the 1st year WMAP data, the double inflation 
models (hybrid+new~\cite{Kawasaki:2003zv,Yamaguchi:2003fp}, 
smooth hybrid+new~\cite{Yamaguchi:2004tn}) were proposed. 
However, since the power spectrum was obtained by analytical
method in those works, the precise form of the spectrum on scales 
corresponding to transition from one inflation to another 
was not clear, which makes it difficult to compare the theoretical
prediction with observations. 

In this paper, therefore, we consider a double inflation model
and calculate the power spectrum by numerical integration of 
evolution equations for density fluctuations. We adopt the smooth hybrid
+new inflation~\cite{Yamaguchi:2004tn} as a double inflation model. 
The model consists of smooth hybrid inflation~\cite{Lazarides:1995vr} and
new inflation~\cite{Kumekawa:1994gx,Izawa:1996dv,Ibe:2006fs}, 
both of which are based on supergravity. 
The running index is
realized by the smooth hybrid inflation. The new inflation 
is necessary because the hybrid inflation with a large running index
only  has small $e$-folds $N \sim 10$ and the remaining
$e$-folds for successful inflation are provided by the new inflation. 
We show that the model can give an appropriate power spectrum which are
consistent with WMAP result. Moreover, we find that large 
density fluctuations are produced through parametric 
resonance~\cite{Kofman:1994rk,Shtanov:1994ce,Kofman:1997yn}
of the inflaton fields at the transition epoch, which
leads to a sharp peak on small scales in the spectrum and formation of
primordial black holes (PBHs). 

This paper is organized as follows: In Sec.II, we introduce the smooth 
hybrid new inflation.
The results of our numerical calculation on the power spectrum of 
the density perturbation are shown in Sec.III.
In Sec.IV, we investigate the parametric resonance in this 
inflationary model which results in a sharp peak found in Sec.III.
In Sec.V, PBH formation is discussed as a consequence of a resonant peak.
Sec.VI is devoted to a discussion. 

In this paper, we set the reduced Planck scale $M_{G}=2.4\times10^{18}{\mathrm{GeV}}$ 
to be unity unless otherwise stated.

%%%%%%%%%%%%%%%%%%%%%%%%%%%%%%%%%%%%%%%%%%%%%%%%%%%%%%%%%
\section{Smooth hybrid new inflation model}
\label{sec:double-inflation}
%%%%%%%%%%%%%%%%%%%%%%%%%%%%%%%%%%%%%%%%%%%%%%%%%%%%%%%%

In this section, we briefly review the smooth hybrid new inflation
model \cite{Yamaguchi:2004tn}, which is based on supergravity.
Its superpotential is given by
\begin{eqnarray}
	W = W_H + W_N,
\end{eqnarray}
where $W_H$ ($W_N$)  is the superpotential responsible for smooth 
hybrid (new) inflation. $W_H$ is written as
\begin{eqnarray}
	W_H &=& S \left( -\mu^2 
	+ \frac{ \left( \Psi \bar{\Psi} \right)^m }{ M^{2(m-1)} } \right).
\end{eqnarray}
Here, $\Psi$ and $\bar{\Psi}$ are a conjugate pair of superfields 
transforming as nontrivial representations of some gauge group $G$.
$S$ is a superfield whose scalar component is the inflaton and
transforms as a singlet under $G$. Moreover, $W_H$ has two symmetries:
one is an $R$-symmetry under which $S\rightarrow e^{2\alpha i}, \Psi \rightarrow 
\Psi$ and $\bar{\Psi}\rightarrow \bar{\Psi}$, and the other is 
a discrete $Z_m$ symmetry under which the combination 
$\Psi \bar{\Psi}$ has unit charge. $M$ sets a cutoff scale which controls 
nonrenormalizable terms in $W_H$.
We include possible coupling constants in the definition of $M$.
$\mu$ sets the scale of the smooth hybrid inflation.

The superpotential $W_N$ is given by
\begin{eqnarray}
	W_N &=& v^2 \Phi - \frac{g}{n+1} \Phi^{n+1}.
\end{eqnarray}
$\Phi$ is the inflaton superfield 
for the new inflation and has a discrete $R$-symmetry, 
$Z_{2nR}$~\cite{Izawa:1996dv}. 
$v$, which satisfies $v \ll \mu$, is the scale of the new inflation, and 
$g$ is a coupling constant of nonrenormalizable terms in $W_N$. 

The K\"ahler potential is given by
\begin{eqnarray}
	K = K_H + K_N,
\end{eqnarray}
\begin{eqnarray}
	K_H &=& |S|^2 + |\Psi|^2 + |\bar{\Psi}|^2, \\
	K_N &=& |\Phi|^2 + \frac{C_N}{4} |\Phi|^4,
\end{eqnarray}
where $C_N$ is a constant smaller than unity.

From $W$ and $K$, we can derive the scalar potential.
We assume $D$-flatness, which coincides with the steepest descent 
direction in the $F$-term contribution, and consider only $F$-term 
contribution.
The scalar potential is given by
\begin{eqnarray}
    V &=& \exp \left[ |S|^2 + |\Psi|^2 + |\bar{\Psi}|^2 + |\Phi|^2 + \frac{C_N}{4}|\Phi|^4 \right] 
    \nonumber \\
    &~& \times \left[ \left| (1+|S|^2) \left( -\mu^2 + \frac{(\bar{\Psi}\Psi)^m}{M^{2(m-1)}} \right) 
    + S^* \Phi \left( v^2 - \frac{g}{n+1} \Phi^n \right) \right|^2 \right. \nonumber \\
    &~& ~~~~ + \left| m\bar{\Psi} \frac{S (\bar{\Psi}\Psi)^{m-1}}{M^{2(m-1)}} + S\Psi^* \left( -\mu^2 
   + \frac{(\bar{\Psi}\Psi)^m}{M^{2(m-1)}} \right) 
   + \Psi^* \Phi \left( v^2 - \frac{g}{n+1}\Phi^n \right) \right|^2 \nonumber \\
   &~& ~~~~ + \left| m\Psi \frac{S (\bar{\Psi}\Psi)^{m-1}}{M^{2(m-1)}} + S\bar{\Psi}^* \left( -\mu^2 
   + \frac{(\bar{\Psi}\Psi)^m}{M^{2(m-1)}} \right) 
   + \bar{\Psi}^* \Phi \left( v^2 - \frac{g}{n+1}\Phi^n \right) \right|^2 \nonumber \\
   &~& ~~~~ + \frac{1}{1 + C_N|\Phi|^2} \left|v^2\left( 1 + |\Phi|^2 
   + \frac{C_N}{2} |\Phi|^4 \right) 
   - \left( 1 + \frac{|\Phi|^2}{n+1} + \frac{C_N |\Phi|^4}{2(n+1)} \right) g |\Phi|^n \right. 
   \nonumber \\
   &~& ~~~~~~~~ \left. \left. + S \Phi^* \left( 1 + \frac{C_N}{2}|\Phi|^2 \right) 
   \left( -\mu^2 + \frac{(\bar{\Psi}\Psi)^m}{M^{2(m-1)}} \right) \right|^2 - 3|W|^2 \right].
\end{eqnarray}
Here, the scalar components of the superfields are denoted 
by the same symbols as the corresponding superfields.
Performing adequate transformations allowed by the symmetries, the complex 
scalar fields are changed into real scalar fields:
\begin{eqnarray}
	\sigma \equiv \sqrt{2} {\mathrm{Re}}S,~
	\psi \equiv 2 {\mathrm{Re}}\Psi = 2 {\mathrm{Re}}\bar{\Psi},~
	\phi \equiv \sqrt{2} {\mathrm{Re}}\Phi.
\end{eqnarray}
In what follows, we will use these real scalar fields.

%%%%%%%%%%%%%%%%%%%%%%%%%%%%%%%%%%%%%%%%
\subsection{Smooth hybrid inflation}
%%%%%%%%%%%%%%%%%%%%%%%%%%%%%%%%%%%%%%%%

First, we make the assumption that initially $|\sigma|$ is
sufficiently large though $|\sigma|<1$ is satisfied, and that $\psi$ and $\phi$ are set around local
minimum.  Indeed, if $|\sigma|$ is sufficiently large, $\psi$ and $\phi$
have effective masses larger than the Hubble parameter $H$, 
so that they roll down to their respective minima quickly. 
Thus, the effective potential of $\sigma$ determines not only 
the dynamics of the smooth hybrid inflation 
but also primordial density fluctuations generated during the
smooth hybrid inflation \cite{Yamaguchi:2005}. Retaining only relevant
terms, it is given by
\begin{eqnarray}
\label{effective_VH}
	V_{H\mathrm{eff}}(\sigma) \simeq 
	\mu^4 \left(
		1 - \frac{2}{27}\frac{M^2 \mu^2}{\sigma ^4} 
		+ \frac{\sigma^4}{8} + \cdots
	\right).
\end{eqnarray}
Hereafter, we consider only the case with $m=2$.
As long as $\sqrt{M\mu} \ll |\sigma| \ll 1$, the effective potential is dominated 
by the false vacuum energy $\mu^4$, and hence inflation takes place.
The Hubble parameter $H$ is almost constant 
$H \simeq H_H \equiv \mu^2/\sqrt{3}$. 
The second term of Eq.(\ref{effective_VH}), which originates from 
nonrenormalizable term of superpotential, has a negative curvature.
On the other hand, the third term of Eq.(\ref{effective_VH}), 
which comes from supergravity correction, has a positive curvature.
Then, for modes crossing the horizon while the third term dominates the dynamics,
the spectrum of curvature perturbation has a spectral index $n_s>1$.
In opposition, for modes crossing the horizon while the second 
term dominates, the spectrum of curvature perturbation has a 
spectral index $n_s<1$. 
In this way, this smooth hybrid inflation model can generate 
the spectrum with the running spectral index suggested by WMAP results.

We estimate the amplitude of curvature perturbation 
${\mathcal{R}}$, spectral index $n_s$ and its running $dn_s/d\ln k$, 
up to second order of slow-roll parameters (Ref.~\cite{Yamaguchi:2004tn}).
According to WMAP three year result, these cosmological parameters are constrained by
\begin{eqnarray}
	\label{eq:WMAP-constraint}
	{\mathcal{R}}^2(k_0) =  24^{+1}_{-2} \times10^{-10},
	~ n_s(k_0) = 1.050^{+0.054}_{-0.072},
	~\frac{dn_s}{d\ln k} = -0.055^{+0.029}_{-0.035}
\end{eqnarray}
at 1$\sigma$ level, where $k_0=0.002 [{\mathrm{Mpc}}^{-1}]$.
We can determine the parameters requiring that the spectrum satisfies these constraints.
However, as shown in Ref.~\cite{Yamaguchi:2004tn}, the number of 
$e$-foldings $N_H$ after the mode with 
$k_0=0.002 [{\mathrm{Mpc}}^{-1}]$ crosses the horizon is estimated 
to be $N_H \le 10$.
This is too small to solve the horizon and the flatness problems, 
and hence the second inflation is needed.

In addition, we must require that $N_H$ is sufficiently large: $N_H \gtrsim 10$,
for, as we will see later, the perturbations produced by the second inflation 
has a very large amplitude. The scale of those perturbations should be
sufficiently small, say $k \gtrsim 1[\mathrm{Mpc^{-1}}]$, in order not 
to conflict with observations. 
It is difficult to meet this request while reproducing the best-fit 
values of WMAP three year result, because the latter requires 
large running.
If we allow 1$\sigma$ range of WMAP three year constraint, we 
can reproduce such a spectrum. We choose the following parameters:
\begin{eqnarray}
\label{eq:parameters_SH}
	\mu=2.04\times10^{-3},~M=1.17.
\end{eqnarray}
This reproduces best-fit values of ${\mathcal{R}}$ and $n_s$, 
but $dn_s/d\ln k$ within the $1\sigma$ range.
\begin{eqnarray}
	{\mathcal{R}}(k_0)=4.9\times10^{-5},~
	n_s(k_0)=1.053,~
	\frac{dn_s}{d\ln k}=-0.032
\end{eqnarray}
at $k=0.002[\mathrm{Mpc^{-1}}]$,
as the result of numerical calculation described in Sec.III.

Note that the present model produces negligibly small tensor modes.
In fact, the tensor to scalar ratio $r$ is estimated by a slow-roll parameter as 
$r=16\epsilon_H < 10^{-3}$.

%%%%%%%%%%%%%%%%%%%%%%%%%%%%%%%%%%%%%%%
\subsection{Oscillatory phase}
%%%%%%%%%%%%%%%%%%%%%%%%%%%%%%%%%%%%%%%

After the smooth hybrid inflation, the oscillatory regime sets in.
$\sigma$ and $\psi$ oscillate around their respective minima,
$\sigma_{\mathrm{min}}=0$ and $\psi_{\mathrm{min}}=2\sqrt{\mu M}$.
Hereafter, we replace $\psi \to \psi_{\mathrm{min}} + \psi$.
Around these minima, their effective masses $m_\sigma$ and $m_\psi$ are given by 
\begin{eqnarray}
\label{eq:mass_of_sigma_and _psi}
	m_\sigma = \sqrt{ \frac{8\mu^3}{M} },~~~~
	m_\psi = \sqrt{ \frac{8\mu^3}{M} + 16\mu^4}.
\end{eqnarray}
These scalar fields $\sigma$ and $\psi$ undergo damped oscillations, whose amplitude 
decreases asymptotically in proportion to $t^{-1} \propto a^{-3/2}$. 
The total energy of oscillating fields $\sigma$ and $\psi$ decreases as $a^{-3}$,  
like non-relativistic matter.
Eventually, contribution of $\phi$, which is about $v^4$, 
dominates the total energy, and the new inflation starts.
The duration of oscillatory phase in terms of the scale factor can be estimated as
\begin{eqnarray}
	\ln \frac{a_{\mathrm{ini}}}{a_c} \simeq \frac{4}{3}\ln\frac{\mu}{v}.
\end{eqnarray}
The subscript $``c"$ indicates that the value is evaluated 
at the end of smooth hybrid inflation, and the subscript $``{\mathrm{ini}}"$ 
indicates that the value is evaluated at the beginning of the new 
inflation.

On the other hand, due to interaction terms between $\sigma,\psi$ 
and $\phi$, $\phi$ also has an effective mass $m_\phi$.
Averaged for a time-scale sufficiently longer than the period of 
oscillation of $\sigma$ and $\psi$, one can estimate $m_\phi$ as
\begin{eqnarray}
	m_\phi^2 \simeq \frac{3}{2} H^2.
\end{eqnarray}
The effective mass of $\phi$ is larger than $H$, so it shows oscillatory 
behavior with a very long period \cite{Kawasaki:1998vx}.
The amplitude of this oscillation decays in proportion to 
$t^{-1/2} \propto a^{-3/4}$.
At the beginning of this phase, $\phi$ can be estimated by the minimum 
value at the end of smooth hybrid inflation, $-(v^2/\mu^2)\sigma_c$,
where $\sigma_c$ is the value of $\sigma$ at the end of smooth hybrid inflation.
In our case, $\sigma_c \sim 0.14$. 
This gives the estimation of mean initial value $\phi_{\mathrm{ini}}$ at the onset of the new inflation
\begin{eqnarray}
    \label{eq:phi_initial}
	\phi_{\mathrm{ini}} \sim -\frac{v^3}{\mu^3} \sigma_c.
\end{eqnarray}

With the total energy density at the reheating $\rho_{\mathrm{reh}}$, 
and assuming that ordinary thermal history after reheating, 
we can estimate the total $e$-foldings of inflation after the horizon crossing of 
the present Hubble scale:
\begin{eqnarray}
\label{eq:e-fold_total}
	N_{\mathrm{tot}}
	\simeq 67 + \frac{2}{3} \ln \frac{\mu}{M_{G}} 
	+ \frac{1}{3} \ln \frac{\rho_{\mathrm{reh}}^\frac{1}{4}}{M_{G}}.
\end{eqnarray}
%%

%%%%%%%%%%%%%%%%%%%%%%%%%%%%%%%%%%%%%
\subsection{New inflation}
%%%%%%%%%%%%%%%%%%%%%%%%%%%%%%%%%%%%%

During the new inflation, the dynamics of $\phi$ is controlled 
by
\begin{eqnarray}
	\frac{dV_N}{d\phi} \simeq -C_N v^4 \phi 
	- 2^\frac{n-2}{2} n g v^{n-1} + 2^{1-n} n g^2 \phi^{2n-1}.
\end{eqnarray}

For $v \ll \mu$ and $\mu$ given by (\ref{eq:parameters_SH}), 
the number of $e$-foldings of the new inflation $N_N$ is estimated 
from Eq.(\ref{eq:e-fold_total})\footnote{%%begin of footnote
In principle, we must determine the decay rate of $\phi$ into other light particles, 
including the standard model particles.
Here, since it is enough to have rough estimation,
we assume gravitationally suppressed interactions,
as was done in \cite{Izawa:1996dv}.} %% end of footnote
as $N_N \lesssim 50$.
Assuming $n=4$, 
we choose $n=4,~C_N=0.04,~v=5.0\times10^{-4},~g=2.0\times10^{-5}$
according to the dynamics of $\phi$, in order to give $N_N \lesssim 50$.

In this new inflation, the scalar potential has a negative minimum:
\begin{eqnarray}
	|\phi| \simeq \sqrt{2} \left( \frac{v^2}{g} \right)^\frac{1}{n},
	~V_{N\mathrm{min}} \simeq -6\left( \frac{n}{n+1} \right)^2 v^4 
	\phi_{\mathrm{min}}^2.
\end{eqnarray}
This gives a negative cosmological constant after inflation.
If we assume that there is another sector which breaks supersymmetry
and that this negative energy density is canceled by a positive 
contribution from supersymmetry breaking, the scale $v$ is related 
to the gravitino mass $m_{3/2}$ as 
\begin{eqnarray}
	m_{3/2} \simeq \frac{n}{n+1} 
	\left( \frac{v^2}{g} \right)^\frac{1}{n}v^2.
\end{eqnarray}
Note that the present model $v \sim 10^{-4}$ gives unacceptably 
large gravitino mass,
$m_{3/2}=1.2\times10^8{\mathrm{TeV}}$.
However, we can abandon the relation between inflation scale $v$ 
and the gravitino mass $m_{3/2}$ by  assuming that the negative 
potential energy $V_{N\mathrm{min}}$ is cancelled not only 
by the contribution from SUSY breaking, but a constant term in the superpotential of another 
sector.
In this paper, we assume that this is the case.

We can estimate analytically the amplitude of primordial curvature 
perturbation for the mode crossing the horizon at the onset of 
the new inflation as
\begin{eqnarray}
	{\mathcal{R}} = \frac{v^2}{2\sqrt{3} \pi \phi_{\mathrm{ini}}} = \frac{\mu^2}{2\sqrt{3} \pi}
	\frac{\mu}{v}\sigma_c^{-1}.
\end{eqnarray}
This is larger than the amplitude at larger scales.
Note that the slow-roll parameter $\eta_N$ is given by
\begin{eqnarray}
	\eta_N = -C_N - 
	2^\frac{2-n}{2} n (n-1) g \frac{\phi^{n-2}}{v^2} \simeq -C_N.
\end{eqnarray}
Therefore, for the cosmologically relevant scales which cross 
the horizon during the new inflation, the spectrum of 
curvature perturbation has an almost constant spectral index
$n_s = 1 -2C_N < 1$.

%%%%%%%%%%%%%%%%%%%%%%%%%%%%%%%%%%%%%%%%
\section{Numerical Calculations}
%%%%%%%%%%%%%%%%%%%%%%%%%%%%%%%%%%%%%%%%%

We now solve numerically the evolution of fluctuations of the scalar fields
$\sigma,\psi$ and $\phi$ until the end of new inflation,
and calculate the spectrum of curvature perturbation after the inflation.
We use scalar potential approximated as
\begin{eqnarray}
\label{eq:potential_cal}
	V &=& V_H + V_N + V_I, \\
\label{eq:potential_cal_H}
	V_H &=&
	\left(
		-\mu^2 + \frac{\psi^{2m}}{4^mM^{2(m-1)}}
	\right)^2
	\left(
		1+\frac{\sigma^4}{8}+\frac{\psi^2}{2}
	\right)
	\nonumber \\
	&~&~~~~~~
	 + \frac{m^2 \sigma^2 \psi^2}{4^{2m-1}} 
	 \left( 
	 	\frac{\psi}{M}
	\right)^{4(m-1)}
 	 + \frac{m \psi^{2m} \sigma^2}{2^{2m-1} M^{2(m-1)}}
	 \left(
		-\mu^2 + \frac{\psi^{2m}}{4^mM^{2(m-1)}}
	\right),	
	 \\ \nonumber \\
\label{eq:potential_cal_N}
	V_N &=& 
	\left( 
		v^2 - \frac{g}{2^\frac{n}{2}} \phi^n 
	\right)^2 
	- \frac{C_N}{2}v^4\phi^2, 
	\\ \nonumber \\
\label{eq:potential_cal_I}
	V_I &=& 
	\left(
		-\mu^2 + \frac{\psi^{2m}}{4^mM^{2(m-1)}} 
	\right)^2 
	\frac{\phi^2}{2}
	 - \left(
	 	-\mu^2+\frac{\psi^{2m}}{4^mM^{2(m-1)}}
	\right)v^2\sigma\phi,
\end{eqnarray}
with the following parameter set:
\begin{eqnarray}
	\mu = 2.04\times10^{-3}&, &~M=1.17,~~v=4.7\times10^{-4} \nonumber \\
	m=2 &,&~n=4,~~C_N=0.04. \nonumber
\end{eqnarray}
We adopt the linear perturbation formalism presented in \cite{Salopek:1989qh}.
We solve the evolution of perturbation in longitudinal gauge, in which
perturbed metric is given by
\begin{eqnarray}
	ds^2 = - (1 + 2\Phi_A) dt^2 + a^2 (1 + 2\Phi_H)\delta_{ij} dx^i dx^j.
\end{eqnarray}
Evolution equations are given by
\begin{eqnarray}
	( \delta\varphi)^{\cdot\cdot}_i+(3H+\Gamma_i)(\delta\varphi)^\cdot_i
	+\frac{k^2}{a^2}\delta\varphi_i
	+ \sum_j\frac{\partial^2 V}{\partial\varphi_i\partial\varphi_j}\delta\varphi_j
	- \left( 2\frac{\partial V}{\partial\varphi_i} 
	+ \dot{\varphi}_i\Gamma_i \right)\Phi_H + 
	4\dot{\varphi}_i\dot{\Phi}_H=0, \nonumber \\
	\label{eq:field-perturbation_evolution}
\end{eqnarray}
\begin{eqnarray}
	\ddot{\Phi}_H+5H\dot{\Phi}_H + \left(\frac{k^2}{3a^2} + \frac{4V}{3}\right)\Phi_H
	- \frac{1}{3}\sum_i\left( 2\frac{\partial V}{\partial\varphi_i}\delta\varphi_i
	 - \dot{\varphi}_i (\delta\varphi)^\cdot_i \right)=0,
	 \label{eq:metric-perturbation_evolution}
\end{eqnarray}
where roman subscripts run over $1,2,3$, and $\varphi_1$, $\varphi_2$, $\varphi_3$ stand for
$\sigma$, $\psi$, $\phi$, respectively.
$\delta\varphi_i$ means fluctuation of corresponding scalar fields.

We calculate curvature perturbation on uniform-density hypersurface $\zeta$:
\begin{eqnarray}
	\zeta = \frac{2}{3} \frac{\rho}{p + \rho}(\Phi_H + H^{-1}\dot{\Phi}_H) +
	\left( 1 + \frac{2 k^2}{9 a^2 H^2}\frac{\rho}{p + \rho} \right) \Phi_H,
\end{eqnarray}
which is constant and equal to comoving curvature perturbation $\mathcal{R}$ 
on superhorizon scales.
Here, $\rho$ and $p$ are total energy density and pressure of background, respectively.
We introduce decay of scalar fields into radiation, with homogeneous and
time-independent decay rates in the manner introduced by \cite{Salopek:1989qh}.
This is done by adding an extra friction terms to the equation of motion of scalar fields.
We treat radiation as a thermal bath, which has contribution to energy density and
pressure of background.
The energy density of radiation $\rho_{\mathrm{r}}$ decays in proportion to $a^{-4}$,
while energy is injected from decay of scalar fields:
\begin{eqnarray}
	\frac{1}{a^4}\frac{d(a^4\rho_{\mathrm{r}})}{dt} =
	\Gamma_\sigma \dot{\sigma}^2 + \Gamma_\psi \dot{\psi}^2 +
	\Gamma_\phi \dot{\phi}^2.
\end{eqnarray}
We set decay rates $\Gamma_\sigma$ and $\Gamma_\psi$ to 
being negligibly small during the whole calculation
: $\Gamma_\sigma = \Gamma_\psi = 10^{-10} M_{G} \ll H_N \ll H_H$.
On the other hand, since $\Gamma_\phi$ gives reheating temperature,
$\Gamma_\phi$ should be far smaller than this.
In our calculations, we regard $\Gamma_\phi$ to be completely negligible:
$\Gamma_\phi=0$.

\begin{figure}[htbp]
	\begin{center}
		\includegraphics[width=.8\linewidth]{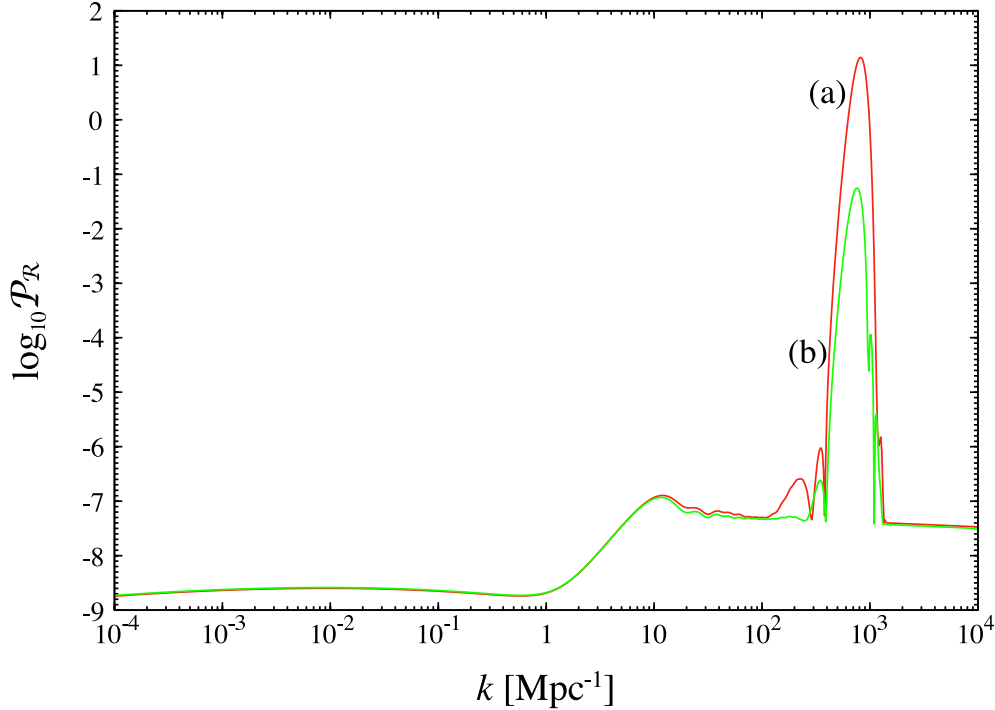}
		\caption{(a) The spectrum of primordial curvature perturbation 
		produced by smooth hybrid new inflation, 
		calculated under the parameter choice (\ref{eq:parameter-LSS})
		and negligibly small decay rate 
		$\Gamma=10^{-10} M_{G}$ of inflaton fields $\sigma$ and $\psi$.
		The largest peak is located at $k=810[{\mathrm{Mpc}}^{-1}]$. 
		(b) The spectrum calculated under the same parameter choice for (a),
		but larger decay rate $\Gamma = 6\times10^{-8}M_{G}$.}
		\label{fig:spectrum}
	\end{center}
\end{figure}

In Fig.\ref{fig:spectrum}-(a), we show the spectrum of primordial 
curvature perturbation calculated numerically.
At $k_0=0.002[{\mathrm{Mpc}}^{-1}]$, this spectrum has
\begin{eqnarray}
	{\mathcal{R}} = 4.9\times10^{-5},~~~n_s=1.053,
	~~~\frac{dn_s}{d\ln k}=-0.032.
	\label{eq:parameter-LSS}
\end{eqnarray}
This is not the best-fit value of the WMAP three year result, 
but within $1\sigma$ range.

%%%%%%%%%%% Figure %%%%%%%%%%%%%%%%%%%%%%%%
%%%%%%%%%%%%%%%%%%%%%%%%%%%%%%%%%%%%%%%%%%%%%

On large scales, the spectrum has a desired shape and amplitude, 
although it does not agree with the best-fit parameters.
The amplitude becomes large at $k=1[{\mathrm{Mpc}}^{-1}]$, 
and is connected to the spectrum produced by the new inflation, 
which has larger amplitude at $k \sim 10[{\mathrm{Mpc}}^{-1}]$.
According to large scale structure observations, the spectrum is 
constrained to be almost flat on scales larger than 
$k \sim 1[{\mathrm{Mpc}}^{-1}]$.
This is satisfied well by the spectrum in Fig.\ref{fig:spectrum}-(a).

A sequence of sharp peaks is seen between $k_0=200[{\mathrm{Mpc}}^{-1}]$ 
and $k_0=1500[{\mathrm{Mpc}}^{-1}]$.
This originates from parametric resonance, as discussed in the next section.
The height of the peaks is determined by competition between 
parametric resonance and decay of the scalar fields $\sigma$, $\psi$.
When the decay rates are large, the fluctuations ($=$ scalar particles)
decay before the amplitudes are amplified through the 
parametric resonance, which results in suppression of the peaks
in the power spectrum of the curvature perturbation.
This is seen in Fig.\ref{fig:spectrum}-(b) where the decay rates
of $\sigma$ and $\psi$ are assumed to be $6\times10^{-8}M_{G}$. 

In other words,
we design the height of the peaks by choosing appropriate couplings
between $\sigma$($\psi$) and the standard model particles. 

\begin{figure}[htbp]
	\begin{center}
		\includegraphics[width=.8\linewidth]{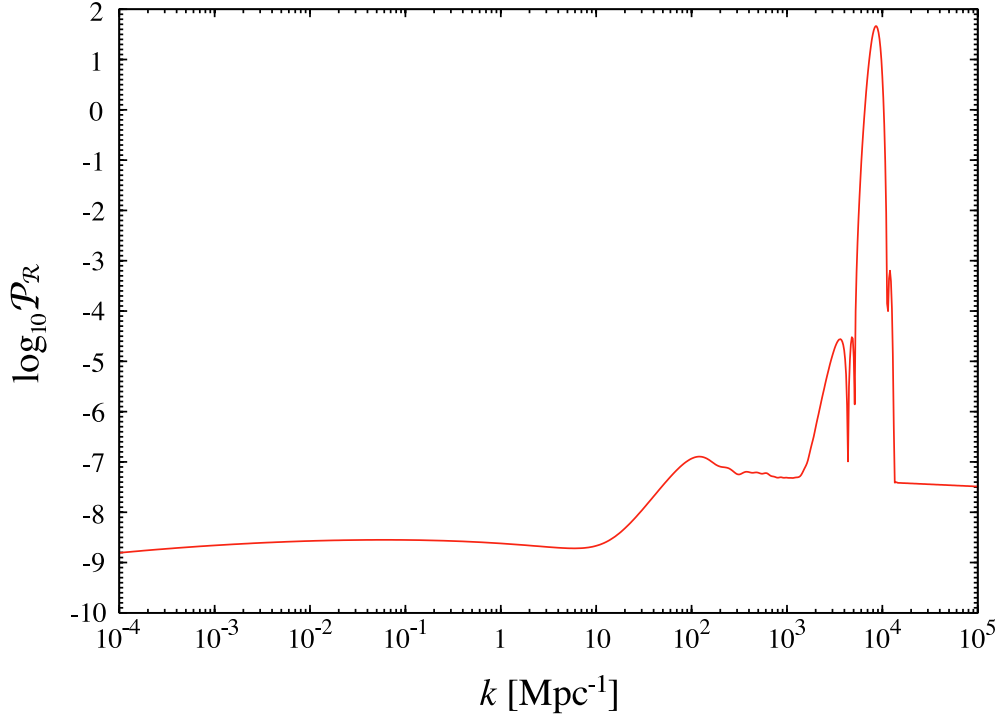}
		\caption{The spectrum of primordial 
		curvature perturbation which is sufficiently small for 
		$k \lesssim 10[\mathrm{Mpc^{-1}}]$. This spectrum is calculated under 
		the parameter choice (\ref{eq:parameter-Lya})
		and negligibly small decay rate $\Gamma = 10^{-10}M_{G}$ 
		of inflaton fields $\sigma$ and $\psi$.}
		\label{fig:spectrum-Lya}
	\end{center}
\end{figure}

If we take observation of Ly-$\alpha$ into account, 
the amplitude of the perturbations should be small
sufficiently for $k \lesssim 10[\mathrm{Mpc^{-1}}]$.
The spectrum shown in Fig.\ref{fig:spectrum-Lya} meets this requirement.
It is given under the parameter set:
\begin{eqnarray}
	\mu = 2.04\times10^{-3}&, &~M=1.12,~~v=5.0\times10^{-4} \nonumber \\
	m=2 &,&~n=4,~~C_N=0.04.
	\label{eq:parameter-Lya}
\end{eqnarray}
At $k_0=0.002[{\mathrm{Mpc}}^{-1}]$, this spectrum has
\begin{eqnarray}
	{\mathcal{R}} = 4.9\times10^{-5},~~~n_s=1.10,
	~~~\frac{dn_s}{d\ln k}=-0.026,
\end{eqnarray}
with $n_s$ and $dn_s/d\ln k$ at the edge of $1\sigma$ 
range of the WMAP three year result.

%%%%%%%%%%%%%%%%% Figure %%%%%%%%%%%%%%%%%%%%%%%%%%%%%%
%%%%%%%%%%%%%%%%%%%%%%%%%%%%%%%%%%%%%%%%%%%%%%%%%%%%%%%%%%%%%%

%%%%%%%%%%%%%%%%%%%%%%%%%%%%%%%%%%%%%%%%%%%%%%%%%%%%%%%%%%%%%%%%%%%%%%
\section{Parametric resonance in smooth hybrid new inflation model}
%%%%%%%%%%%%%%%%%%%%%%%%%%%%%%%%%%%%%%%%%%%%%%%%%%%%%%%%%%%%%%%%%%%%%%

In this section, we will investigate the resonant amplification of 
fluctuations of the scalar fields $\delta\sigma_k$ and $\delta\psi_k$, 
and consequent amplification of $\delta\phi_k$.
This causes a strong peak in the spectrum of primordial curvature 
perturbation.
Since these two regimes have very different time scales, we will describe 
them separately.

%%%%%%%%%%%%%%%%%%%%%%%%%%%%%%%%%%%%%%%%
\subsection{Parametric resonance regime}
%%%%%%%%%%%%%%%%%%%%%%%%%%%%%%%%%%%%%%%%

\begin{figure}[htbp]
	\begin{center}
		\includegraphics[width=.8\linewidth]{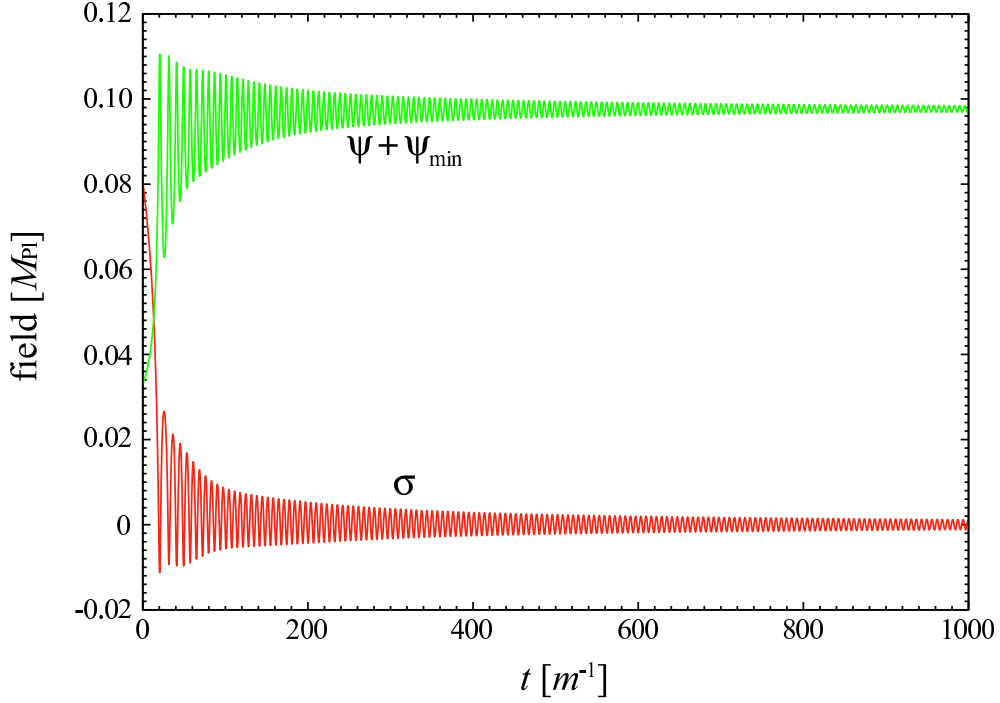}
		\caption{Oscillation of backgrounds $\sigma$ and $\psi$.
		Time variable $t$ is indicated in unit of $m_\sigma^{-1}$, 
		here $m=m_\sigma\simeq m_\psi$. $t=0$ is taken to be 
		$\ln a=-114.5$, at the time shortly before the oscillatory phase begins.}
		\label{fig:BGoscillation}
	\end{center}
\end{figure}

After the smooth hybrid inflation ends, 
$\sigma$ and $\psi$ begin to oscillate about their respective minima.
We show these oscillating backgrounds in Fig.\ref{fig:BGoscillation}.
Hereafter, we normalize the scale factor to be $a=1$ at present.

%%%%%%%%%%%% Figure %%%%%%%%%%%%%%%
%%%%%%%%%%%%%%%%%%%%%%%%%%%%%%%%%%%%

The evolution is very complicated.
Because the cross terms which includes both $\sigma$ and $\psi$ 
exist in the scalar potential, $\delta\sigma_k$ and $\delta\psi_k$ 
contributes to the evolution of each other.
Moreover, at the beginning of the oscillatory phase, 
the second order terms in oscillating backgrounds $\sigma$ and $\psi$ 
contribute to the evolution as well as the first order terms, 
which makes analytical understanding of parametric resonance difficult.

Retaining only the first order terms, and neglecting metric perturbations,
evolution equations of $\delta\sigma_k$ and $\delta\psi_k$ are given by
\begin{eqnarray}
	\label{eq:resonance-sigma}
	(\delta\sigma_k)^{\cdot\cdot} + 3H(\delta\sigma_k)^\cdot + \left[ \frac{k^2}{a^2}
	 + m_\sigma^2 + \frac{24\mu^2}{M^2}\sqrt{\mu M}\psi \right]\delta\sigma_k
	 + \frac{24\mu^2}{M^2}\sqrt{\mu M}\sigma\delta\psi_k = 0, \\
	\label{eq:resonance-psi}
	(\delta\psi_k)^{\cdot\cdot} + 3H(\delta\psi_k)^\cdot + \left[ \frac{k^2}{a^2}
	 + m_\psi^2 + \frac{36\mu^2}{M^2}\sqrt{\mu M}\psi \right]\delta\psi_k
	 + \frac{24\mu^2}{M^2}\sqrt{\mu M}\sigma\delta\sigma_k = 0,
\end{eqnarray}
and retaining lowest order, evolution equations of backgrounds $\sigma$ and $\psi$ are given by
\begin{eqnarray}
	\label{eq:osc-sigma}
	\ddot{\sigma} + 3H\dot{\sigma} + m_\sigma^2\sigma = 0, \\
	\label{eq:osc-psi}
	\ddot{\psi} + 3H\dot{\psi} + m_\psi^2\psi = 0.
\end{eqnarray}

In order to understand their behaviors approximately, we will recast Eqs.(\ref{eq:resonance-sigma}) and (\ref{eq:resonance-psi}) into following form:
\begin{eqnarray}
	\label{eq:mathieu-sigma}
	\delta\tilde{\sigma}'' 
	+ \left[A_\sigma
	- 2q_{\sigma\sigma}\cos\left(2(1+d)z + \Delta \right)\right]\delta\tilde{\sigma} 
	- 2q_{\sigma\psi}\cos(2z)\delta\tilde{\psi}
	= 0, \\ 
	\label{eq:mathieu-psi}
	\delta\tilde{\sigma}''
	+ \left[A_\psi
	- 2q_{\psi\psi}\cos\left(2(1+d)z + \Delta \right)\right]\delta\tilde{\psi} 
	- 2q_{\sigma\psi}\cos(2z)\delta\tilde{\sigma}
	= 0.
\end{eqnarray}
Here, the prime represents the derivative with respect to the variable $z$ defined as $2z = m_\sigma t - \pi/2$. 
$\Delta$ is a possible phase, and $1+d \equiv m_\psi/m_\sigma$, which satisfies $d \ll 1$.
Furthermore, we approximated the oscillating background as
\begin{eqnarray}
	\sigma &\simeq& - \Sigma \sin(m_\sigma t), \\
	\psi &\simeq& - \Psi \sin(m_\psi t + [\mathrm{phase~difference}]).	
\end{eqnarray} 
We also rescaled the perturbations $\delta\sigma_k$ and $\delta\psi_k$ as 
$\delta\tilde{\sigma} \equiv a^{3/2} \delta\sigma_k$ 
and $\delta\tilde{\psi} \equiv a^{3/2} \delta\psi_k$,
and neglected ${\mathcal{O}}(H^2)$ terms.
The coefficients
 $A_\sigma,~A_\psi,~q_{\sigma\sigma},~q_{\sigma\psi},~q_{\psi\sigma}$, and $q_{\psi\psi}$
 are given by
\begin{eqnarray}
	A_\sigma = 4 + \frac{4k^2}{a^2m_\sigma^2},
	~A_\psi = 4 + 4\left( \frac{k^2}{a^2m_\sigma^2} + d \right),
\end{eqnarray}
\begin{eqnarray}
	q_{\sigma\sigma} = \frac{48\mu^2\sqrt{\mu M}}{M^2m_\sigma^2}\Psi,~
	q_{\sigma\psi} = \frac{48\mu^2\sqrt{\mu M}}{M^2m_\sigma^2}\Sigma,~
	q_{\psi\psi} = \frac{72\mu^2\sqrt{\mu M}}{M^2m_\sigma^2}\Psi.
\end{eqnarray}

At the beginning of the oscillatory phase, 
amplitudes are estimated as $\Sigma \sim \Psi \sim 0.01$.
Neglecting $d \ll 1$, we get $A_\sigma \simeq A_\psi \simeq 4\frac{k^2}{a^2m_\sigma^2} +4 \gtrsim 4$, 
$q_{\sigma\sigma} \simeq q_{\sigma\psi} \sim 0.62$,
 and $q_{\psi\psi} \sim 0.93$.
Eqs.(\ref{eq:mathieu-sigma}) and (\ref{eq:mathieu-psi}) look like the Mathieu equation 
\cite{MathieuFunctions},
\begin{eqnarray}
	x'' + [A - 2q\cos(2z)]x = 0.
\end{eqnarray}
Thus, we expect instability solutions.

To confirm the existence of the instability, 
we will show some results of numerical calculations.
First, we have solved the coupled evolution equations (\ref{eq:resonance-sigma}) 
- (\ref{eq:osc-psi}) numerically.
In order to trace evolutions of perturbations $\delta\sigma_k$ and $\delta\psi_k$
and to see the occurrence of resonant amplification,
we neglected expansion of the universe: we put $H=0$.

\begin{figure}[htbp]
	\begin{center}
		\includegraphics[width=.8\linewidth]{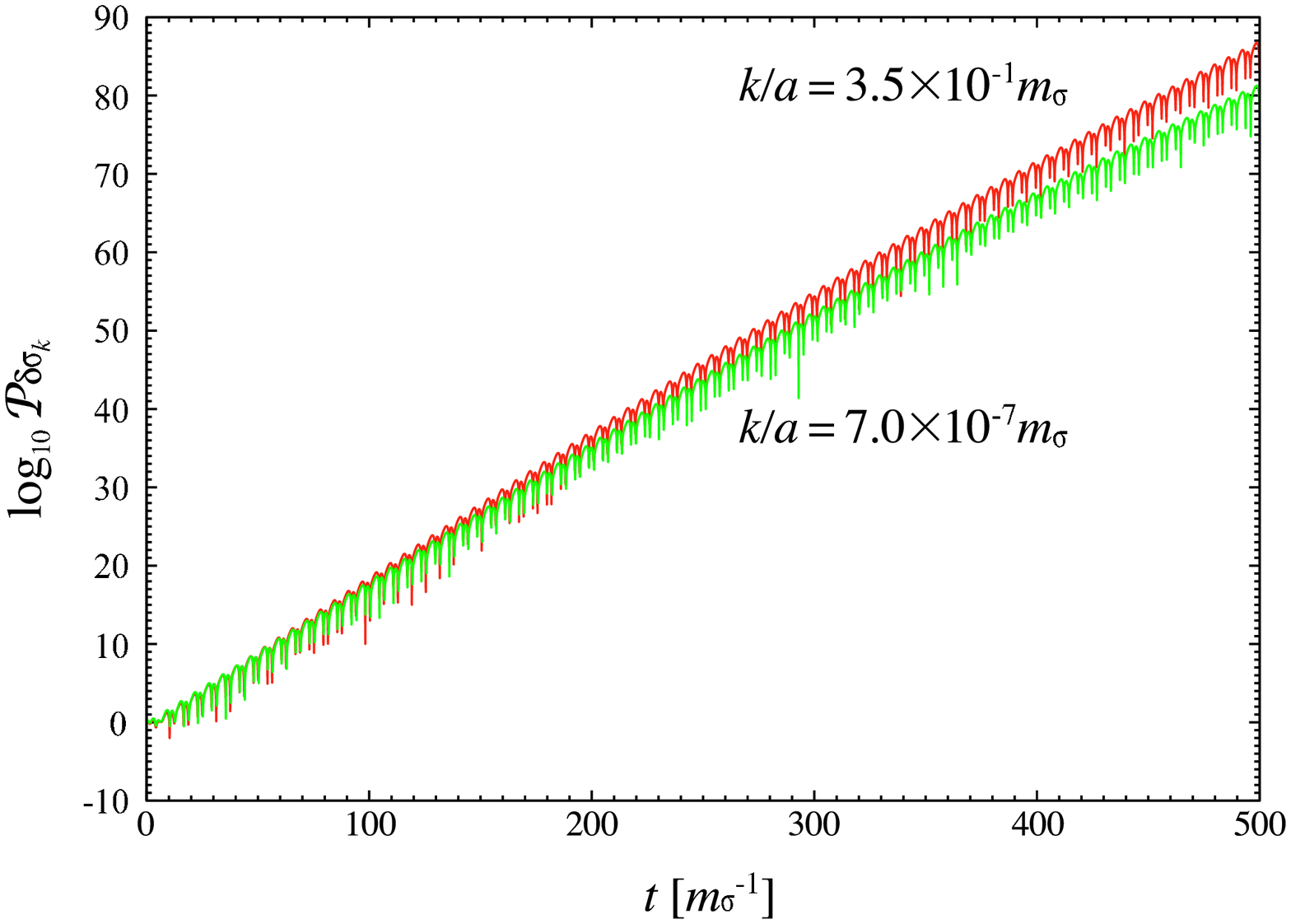}
		\caption{Time evolutions of ${\mathcal{P}}_{\delta\sigma_k}$ under 
		Eqs.(\ref{eq:resonance-sigma}) - (\ref{eq:osc-psi}), 
		for $k/a \simeq 3.5\times10^{-1}m_\sigma$ and 
		$k/a \simeq 7.0\times10^{-7}m_\sigma$.
		}
		\label{fig:time_evolution-approx-sigma}
		\includegraphics[width=.8\linewidth]{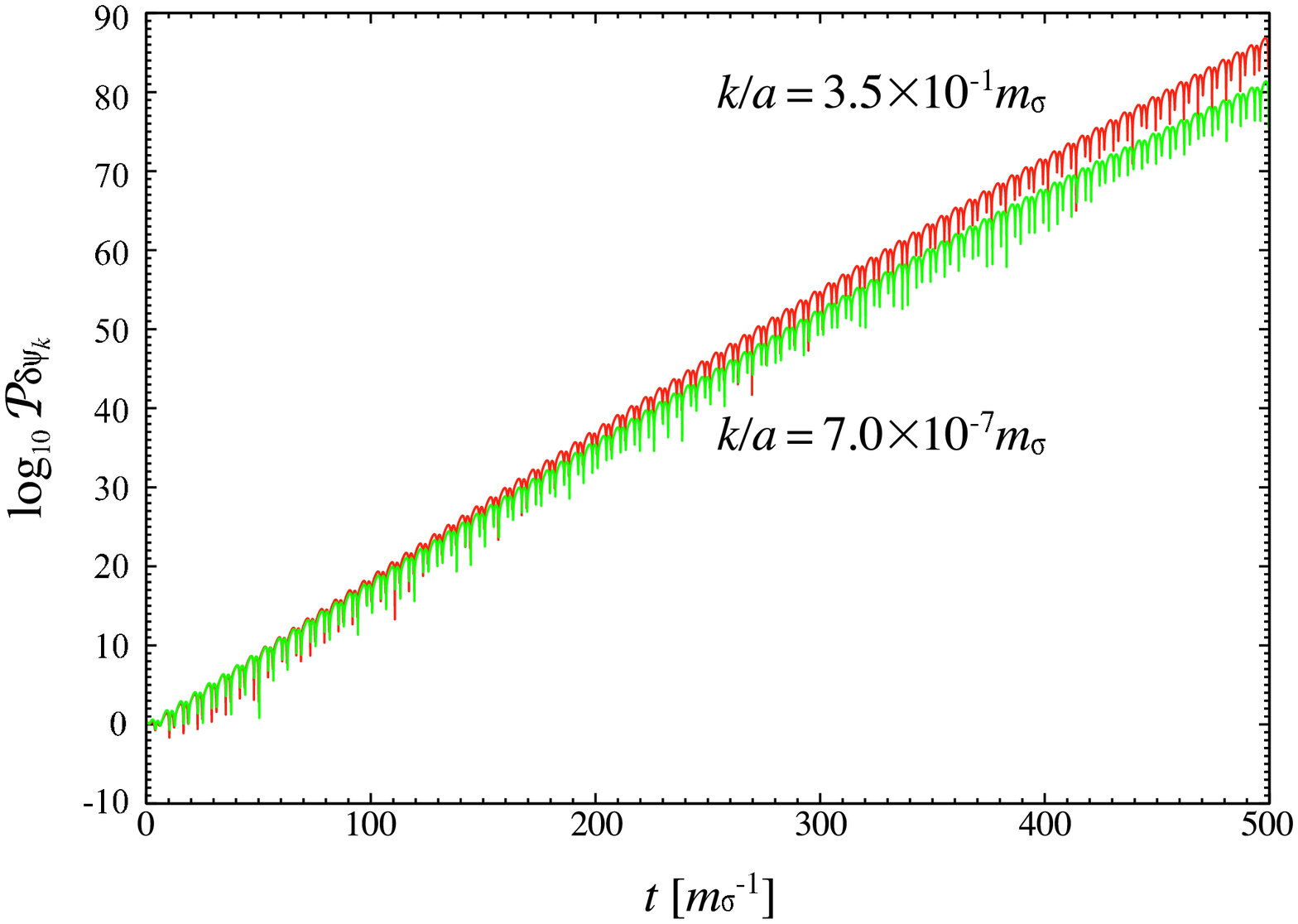}
		\caption{Time evolutions of  ${\mathcal{P}}_{\delta\psi_k}$ under 
		Eqs.(\ref{eq:resonance-sigma}) - (\ref{eq:osc-psi}), 
		for $k/a \simeq 3.5\times10^{-1}m_\sigma$ and 
		$k/a \simeq 7.0\times10^{-7}m_\sigma$.
		}
		\label{fig:time_evolution-approx-psi}
	\end{center}
\end{figure}

Figure \ref{fig:time_evolution-approx-sigma} and Figure \ref{fig:time_evolution-approx-psi} 
show time evolutions of power spectra
\footnote{%%%  Begin of Footnote %%%
For Fourier modes $g_{\vec{k}}$ of a perturbation $g(\vec{x})$, 
the power spectrum ${\mathcal{P}}_g$ is defined by
\[
	\langle g^*_{\vec{k}} g_{\vec{k}'} \rangle \equiv \delta^{(3)} 
	(\vec{k} - \vec{k}') \frac{2\pi^2}{k^3}{\mathcal{P}}_g.
\]
Here, $\langle \cdots \rangle$ represents an ensemble average.
}  %%% End of Footnote  %%%%
of scalar field perturbation 
${\mathcal{P}}_{\delta\sigma_k}$ and ${\mathcal{P}}_{\delta\psi_k}$ under 
Eqs. (\ref{eq:resonance-sigma}) - (\ref{eq:osc-psi}).
We use $\Sigma=1.08\times10^{-2}$ and $\Psi=1.16\times10^{-2}$, 
which are evaluated at $\ln a= -114$
from the full numerical calculation of smooth hybrid new inflation given in Sec.III.
Initial amplitudes $|\delta\sigma_k|$ and $|\delta\psi_k|$ are set to give
$\log {\mathcal{P}}_{\delta\sigma_k} = 0$ and 
$\log {\mathcal{P}}_{\delta\psi_k} = 0$, respectively.
We concentrate on two modes: $k/a \simeq 3.5\times10^{-1}m_\sigma$ and 
$k/a \simeq 7.0\times10^{-7}m_\sigma$.
These modes correspond to 
$k = 1000[{\mathrm{Mpc}}^{-1}]$ and $k = 0.002[{\mathrm{Mpc}}^{-1}]$ at present.
We can see that resonant amplification takes place. 
Rate of exponential amplification varies depending on $k$.
We also find that $\delta\sigma_k$ and $\delta\psi_k$ show
almost identical evolution.

%%%%%%%%%%%%%%%%%%%%%%% Figure %%%%%%%%%%%%%%%%%%%%%
\begin{figure}[htbp]
	\begin{center}
		\includegraphics[width=.8\linewidth]{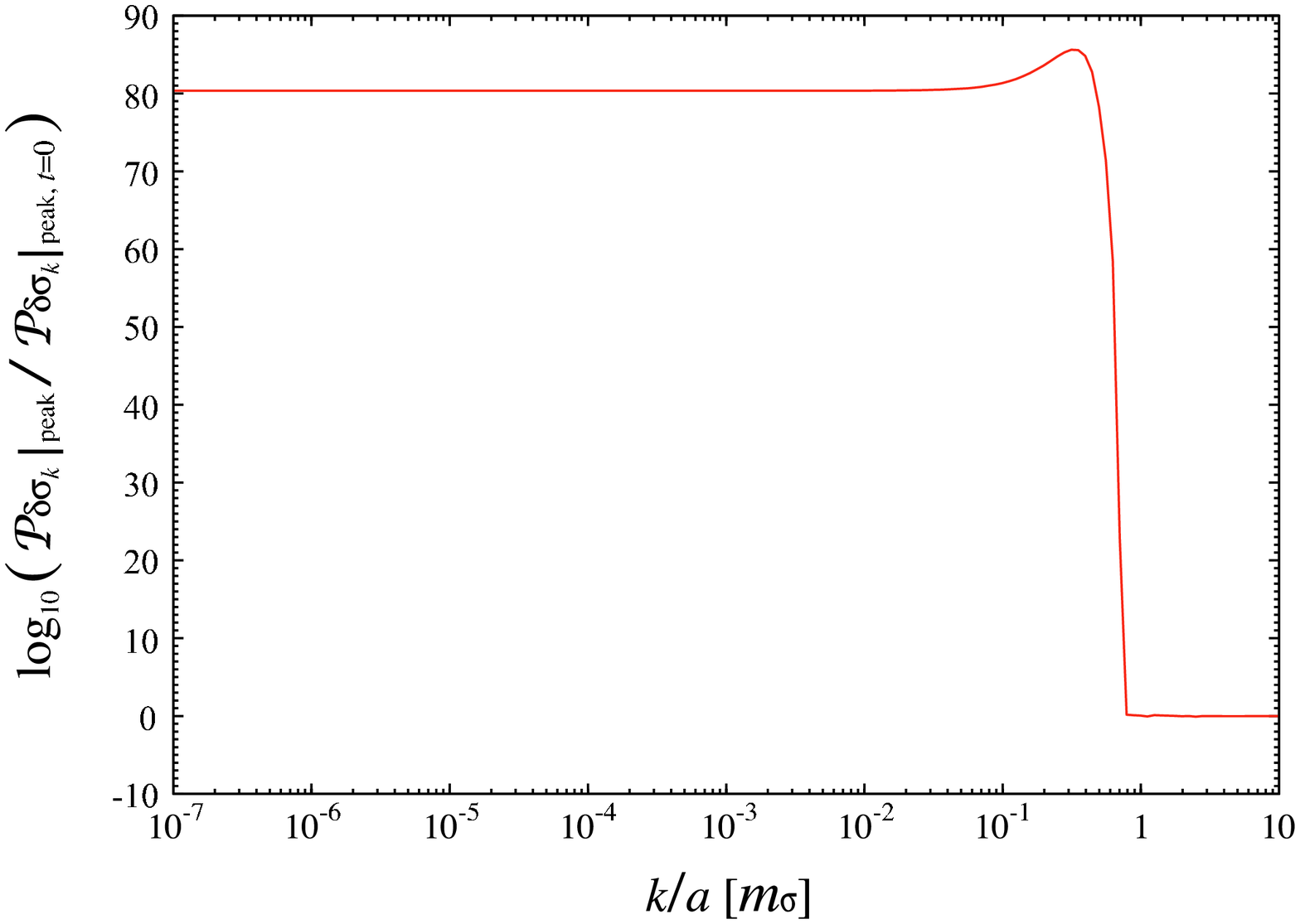}
		\caption{Scale dependence of amplification under the same configuration 
		as that used for Fig.\ref{fig:time_evolution-approx-sigma} and 
		Fig.\ref{fig:time_evolution-approx-psi}. 
		We show $k$-dependence of 
		$\left({\mathcal{P}}_{\delta\sigma_k}|_{\mathrm{peak}}\right) /
		\left({\mathcal{P}}_{\delta\sigma_k}|_{\mathrm{peak,t=0}}\right)$.
		Most efficient amplification takes place at $k/a\simeq3.5\times10^{-1}m_\sigma$, 
		at the time $t=500m_\sigma^{-1}$.
		}
		\label{fig:spectrum-amplification-approx}
	\end{center}
\end{figure}

%%%%%%%%%%%%%%%%%%%%%%%%%%%%%%%%%%%%%%%%%%%%%%%%%%%%%%
Figure \ref{fig:spectrum-amplification-approx} shows scale dependence of 
amplification under the same configuration as that used for Fig.\ref{fig:time_evolution-approx-sigma}.
We indicate only the amplification of $\delta\sigma_k$, at the time $t=500m_\sigma^{-1}$.
Since $\delta\sigma_k$ is oscillating rapidly,
we use ${\mathcal{P}}_{\delta\sigma_k}|_{\mathrm{peak}}$, 
the peak value of oscillating ${\mathcal{P}}_{\delta\sigma_k}$ 
around at some specific moment, 
which is approximately equal to the amplitude of oscillation.
We can see that small momentum modes $k^2/a^2 \ll m_\sigma^2$ show identical amplification.
Most efficient amplification takes place at $k/a \simeq 3.5\times10^{-1}m_\sigma$.

%%%%%%%%%%%%%%%%%%%%%%% Figure %%%%%%%%%%%%%%%%%%%%%
%%%%%%%%%%%%%%%%%%%%%%%%%%%%%%%%%%%%%%%%%%%%%%%%%%%%%%

Next, in order to compare the above discussion with the actual time evolution of 
smooth hybrid new inflation model, 
we numerically calculate the evolution of the fluctuation $\delta\sigma_k$,
in the same way as described in Sec.III,
solving all linear evolution equations of scalar fields 
Eq.(\ref{eq:field-perturbation_evolution}) 
and metric perturbation Eq.(\ref{eq:metric-perturbation_evolution}) 
under the scalar potential $V$ given by Eqs.(\ref{eq:potential_cal}) - (\ref{eq:potential_cal_I}),
in the expanding universe.
 
\begin{figure}[htbp]
	\begin{center}
		\includegraphics[width=.8\linewidth]{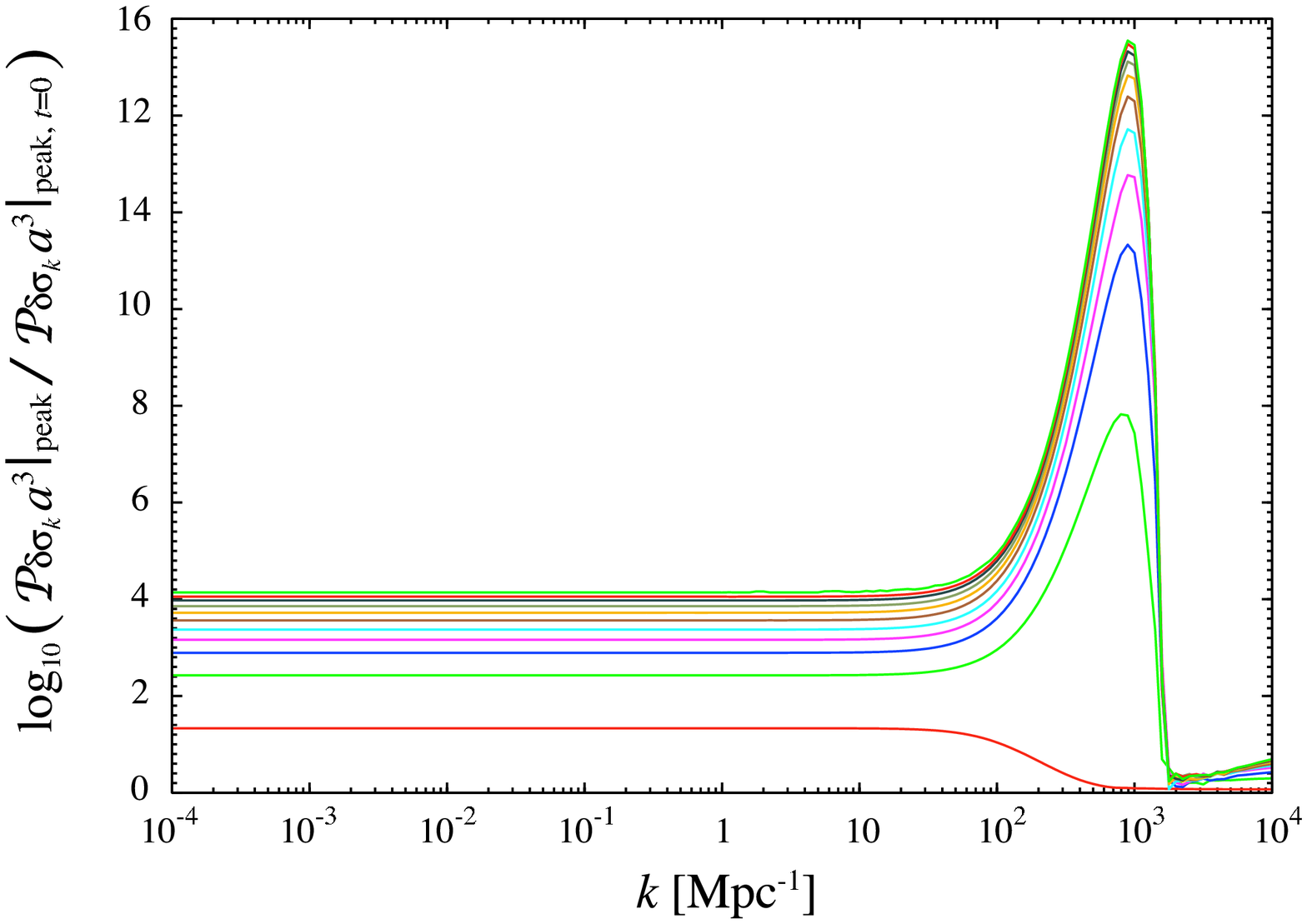}
		\caption{Scale-dependence of the resonant amplification.
		We show $k$-dependence of 
		$\left({\mathcal{P}}_{\delta\sigma_k}a^3|_{\mathrm{peak}}\right) /
		\left({\mathcal{P}}_{\delta\sigma_k}a^3|_{\mathrm{peak,t=0}}\right)$.
		The bottom line indicates $t=15m_\sigma^{-1}$, the rest show various 
		$t$ from $t=100m_\sigma^{-1}$ (next-to-bottom line) to $t=1000m_\sigma^{-1}$ 
		(top line), varying by $100m_\sigma^{-1}$ between each lines.
		At $t=15m_\sigma^{-1}$, amplification by tachyonic instability 
		can be seen: all modes $k/a \ll m$ are equally amplified. 
		After that, resonant peak appears around at
		$k\simeq 1000[{\mathrm{Mpc}}^{-1}]$.}
		\label{fig:spectrum_amplification}
	\end{center}
\end{figure}

Figure \ref{fig:spectrum_amplification} shows the mode dependence 
of efficiency at various times while the resonant amplification takes place.
We use ${\mathcal{P}}_{\delta\sigma_k}|_{\mathrm{peak}}$, for the same reason 
that we did so for Fig.\ref{fig:spectrum-amplification-approx}.
Since all modes of fluctuations of scalar fields decrease 
because of the Hubble friction term,
the amplification is suppressed during the expansion of the universe, 
so that modes around the strong peak are amplified significantly
in contrast to other modes.
At $t=15m_\sigma^{-1}$,  we can see efficient amplification of low-momentum modes.
Since the effective potential of $\sigma$ is tachyonic at the end of hybrid inflation, 
all modes with $k^2/a^2 \ll m_\sigma^2$ are amplified efficiently by tachyonic 
instability \cite{Felder:2000hj,Copeland:2002ku}.
No significant peak appears in the spectrum of scalar perturbation 
at this time.

%%%%%%%%%%%%%%%%%%%%%%% Figure %%%%%%%%%%%%%%%%%%%%%
%%%%%%%%%%%%%%%%%%%%%%%%%%%%%%%%%%%%%%%%%%%%%%%%%%%

%%%%%%%%%%%%%%%%%%%%%%%%%%%%%%%%%%%%%%%%%%%%%%%%%%%
\subsection{Forced oscillation of $\delta\phi_k$}
%%%%%%%%%%%%%%%%%%%%%%%%%%%%%%%%%%%%%%%%%%%%%%%%%%%

Now, we turn to the evolution of $\delta\phi_k$. 
We will focus on subhorizon modes at the beginning of oscillatory phase: 
$k/a > H_H = \mu^2/\sqrt{3}$ with $\ln a=-114.5$.
For superhorizon modes, curvature perturbation is already 
frozen out, and evolution of $\delta\phi_k$ after smooth hybrid 
inflation is irrelevant.

During the smooth hybrid inflation, evolutions of 
$\delta\sigma_k, \delta\psi_k$ and $\delta\phi_k$ are all dominated 
by $k^2/a^2$ term.
They decay like $|\delta\sigma|^2 \propto a^{-2}$ from the same 
initial condition given out of quantum fluctuation.
So, we can estimate that $\delta\sigma_k, \delta\psi_k$ and 
$\delta\phi_k$ have the same amplitude at the beginning of 
the oscillatory phase.

Evolution equations for Fourier modes $\delta\phi_k$ are given by
\begin{eqnarray}
	(\delta\phi_k)^{\cdot\cdot} & + & 3H(\delta\phi_k)^\cdot 
	+ \left[ \frac{k^2}{a^2}
	 + \frac{4\mu^3}{M}\psi^2 \right]\delta\phi_k 
	 - \frac{2\mu}{M}\sqrt{\mu M}v^2
	 \left( \psi\delta\sigma_k + \sigma\delta\psi_k \right) = 0.
    \label{eq:evolution-phi}
\end{eqnarray}
No resonant amplification due to oscillating $\sigma$ and $\psi$ 
occurs, since $(4\mu^3/M)\Psi^2 \ll k^2/a^2 \ll m^2$.

Let us assume sufficiently large amplification of $\delta\sigma_k$ 
or $\delta\psi_k$ occurs so that 
$( \psi\delta\sigma_k + \sigma\delta\psi_k )$ terms dominates 
Eq.(\ref{eq:evolution-phi}). The condition for that is given by
\begin{eqnarray}
\label{eq:condition_forced-oscillation}
	\left| \frac{k^2}{a^2}\delta\phi \right| \ll 
	\left| \frac{2\mu}{M}\sqrt{\mu M}v^2\Psi\delta\sigma_k \right|.
\end{eqnarray}
In our case, this requires more than $10^4$ amplification of 
$\delta\sigma_k$ relative to $\delta\phi_k$.
Since we have more than $10^5$ amplification at the resonant peak, 
this condition is satisfied very well.

Once the $( \psi\delta\sigma_k + \sigma\delta\psi_k )$ term dominates 
Eq.(\ref{eq:evolution-phi}), $\delta\phi_k$ undergoes forced oscillation 
with the source term $( \psi\delta\sigma_k + \sigma\delta\psi_k )$.
One can expect that this results in an amplification of $\delta\phi_k$, 
but there is another subtlety.
In our case, $\psi$ and $\delta\psi_k$ oscillate with frequency 
about $ m_\psi$, while $\sigma$ and $\delta\sigma_k$ oscillate 
with about $m_\sigma$.
Since these two frequencies are slightly different from each other, 
$\psi\delta\sigma_k$ and $\sigma\delta\psi_k$ show log-period 
oscillatory behavior as 
$(C/2)\left[ \cos(m_\sigma \delta t) 
- \cos \left(m_\sigma (2+\delta)t\right) \right]$,
where $C$ is the amplitude.
Then, $\delta\phi_k$ has a solution like
\begin{eqnarray}
	\delta\phi_k \sim -\frac{Ca^2}{2k^2}\cos(m_{\sigma}\delta t).
\end{eqnarray}

\begin{figure}[htbp]
	\begin{center}
		\includegraphics[width=.8\linewidth]{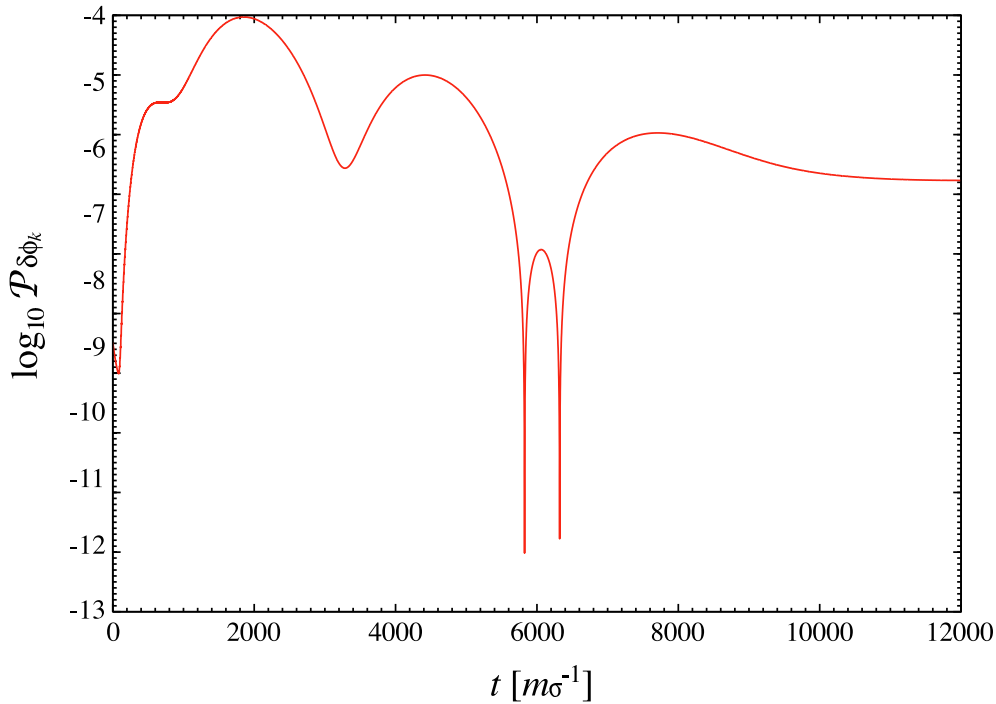}
		\caption{Long-period oscillation of ${\mathcal{P}}_{\delta\phi_k}$ 
		for the mode of resonant peak: $k=810[{\mathrm{Mpc}}^{-1}]$.
		Sharp valley indicates that the oscillation crosses zero.
		The period is estimated to be 
		$\Delta t = 2\pi/\Delta m \simeq 2.6\times10^3 m_\sigma^{-1}$, 
		which agrees well.}
		\label{fig:long-period_oscillation}
	\end{center}
\end{figure}

As a result, $\delta\phi_k$ shows long-period oscillation, 
whose frequency is estimated by $\Delta m$.
This induces large oscillation of the amplitude of $\delta\phi_k$, 
while the mean value of oscillation is still 
decaying.
Meanwhile, the source term decays with 
$| \psi\delta\sigma_k | \propto a^{-3}$, 
since $\sigma$ and $\psi$ behave as massive scalar fields.
On the other hand, $H$ takes a constant value after the beginning 
of the new inflation.
Therefore, contribution from Hubble friction term eventually 
becomes relevant and $\delta\phi_k$ ceases oscillation.
Afterwards, $\delta\phi_k$ decays until the horizon crossing, 
and then it is fixed.
$\delta\phi_k$ at that time determines the amplitude of primordial 
curvature perturbation:
\begin{eqnarray}
	{\mathcal{P_R}} = \left( \frac{H}{\dot{\phi}} \right)^2 
	{\mathcal{P}}_{\delta\phi_k}.
\end{eqnarray}
Since the amplitude of $\delta\phi_k$ is determined by the phase 
of large oscillation, $k$-dependence of $\delta\phi_k$ at horizon 
crossing is oscillatory.
Thus, resultant spectrum of the primordial curvature perturbation 
has a strong peak, ``sliced out" from the single resonant peak 
generated by parametric resonance of $\sigma$ and $\psi$ as shown in
Fig.\ref{fig:long-period_oscillation}, where we show the time evolution 
of $\delta\phi_k$ in this regime.

%%%%%%%%%%%%%%%%%%  Figure %%%%%%%%%%%%%%%%%%%%%%%%%%%%%%
%%%%%%%%%%%%%%%%%%%%%%%%%%%%%%%%%%%%%%%%%%%%%%%%%%%%%%%%%%

Let us reexamine the mechanism which induces this resonant peak by comparing estimation of the strength of peak and the result of numerical calculation.

At the beginning of long-period oscillation, the equation of motion (\ref{eq:evolution-phi}) is dominated by $(k^2/a^2)\delta\phi_k$ term and $(\psi\delta\sigma_k + \sigma\delta\psi_k)$ term.
Therefore we can estimate the amplitude of long-period oscillation of $\delta\phi_k$ from above discussion.
The first peak of ${\mathcal{P}}_{\delta\phi_k}$ is at about $t \approx 1800 m_\sigma^{-1}$, 
$\ln a \approx -111.8$.
At that time, the amplitude of $\delta\sigma_k$ is given by 
${\mathcal{P}}_{\delta\sigma_k} \approx 4.5\times10^2$, 
while the amplitude of oscillating background is $\Psi \approx 3.2\times10^{-4}$, 
according to the numerical calculation.
We can estimate the amplitude of the long-period oscillation of $\delta\phi_k$, in terms of the power spectrum for $k=810[{\mathrm{Mpc}}^{-1}]$,
\begin{eqnarray}
	{\mathcal{P}}_{\delta\hat{\phi}_k} \simeq \frac{a^4}{k^4} 
	\frac{4\mu^3 v^4 \Psi^2}{M}{\mathcal{P}}_{\delta\hat{\sigma}_k} 
	\simeq 1.7\times10^{-5}.
\end{eqnarray}
Hereafter, we will concentrate on amplitudes $\hat{\delta\sigma_k}$ and $\hat{\delta\phi_k}$ of the oscillations of $\delta\sigma_k$ and $\delta\phi_k$, respectively.
We approximated that the $\sigma\delta\psi_k$ term gives the same contribution as that of 
$\psi\delta\sigma_k$ term.
The numerical calculation gives ${\mathcal{P}}_{\delta\hat{\phi}_k} \approx 9.3\times 10^5$.
Our estimation is very rough, but gives the same order of the result of numerical calculation.

At the time the long-period oscillation ceases, 
this estimation underestimates the amplitude ${\mathcal{P}}_{\delta\hat{\phi}_k}$.
At $t \approx 7500 m_\sigma^{-1}$ and $\ln a \approx -108.8$, the long-period oscillation ceases, with 
${\mathcal{P}}_{\delta\hat{\sigma}_k} \approx 3.7\times10^{-2}$ and
$\Psi \approx 3.5\times10^{-6}$.
These give ${\mathcal{P}}_{\delta\hat{\phi}_k} \simeq 9.9\times10^{-8}$,
while the numerical result is ${\mathcal{P}}_{\delta\hat{\phi}_k} \approx 1.0\times10^{-6}$.
This is because of the complexity of the evolution: 
since $k/aH \sim {\mathcal{O}}(1)$, the evolution is in the transition regime 
between subhorizon to superhorizon, 
and hence the evolution of $|\delta\phi_k|$ marginally freezes.

On the other hand, we can verify that $\delta\phi_k$ at the horizon crossing determines 
the curvature perturbation.
At the horizon crossing, numerical calculation gives 
${\mathcal{P}}_{\delta\hat{\phi}_k} \approx 3.0\times10^{-7}$.
At that time, $H \simeq 1.3\times10^{-7}$ and $\dot{\phi}\simeq -1.3\times10^{-11}$.
We can get
\begin{eqnarray}
	{\mathcal{P_R}}\bigl|_{\mathrm{peak}} \lesssim \left( \frac{H}{\dot{\phi}} \right)^2 {\mathcal{P}}_{\delta\hat{\phi}_k} \bigl|_{\mathrm{cross}} \simeq 29.
\end{eqnarray}
This gives $\log{\mathcal{P_R}} \lesssim 1.5$ at the peak, which is the same order
of the result of numerical calculation $\log{\mathcal{P_R}} \approx 1.15$.

%%%%%%%%%%%%%%%%%%%%%%%%%%%%%%%%%%%%%%%
\section{PBH formation from resonant peak}
%%%%%%%%%%%%%%%%%%%%%%%%%%%%%%%%%%%%%%%
The presence of the large sharp peak around $k\simeq 10^3 [{\mathrm{Mpc}}^{-1}]$ 
may lead to interesting consequences, one of which is the formation
of primordial black holes.
From the peak position $k=810[{\mathrm{Mpc}}^{-1}]$ 
of the spectrum shown in Fig.\ref{fig:spectrum}-(a),
the mean mass of resultant PBHs is estimated as 
\begin{eqnarray}
\label{mass_PBH}
	m_{\mathrm{BH}} = 1 \times 10^9 M_{\odot}.
\end{eqnarray}
Here we assume that the whole mass within the overdense region 
when it enters the horizon collapses into a black hole.
The abundance of such a massive PBH is constrained 
by assuming that they must not overclose the universe \cite{Green:1997sz}.
In this case, initial mass fraction $\beta$ is constrained to be $\beta < 4.6\times10^{-6}$.

We estimate the black hole abundance, 
following \cite{Yokoyama:1999xi} based on numerical simulation of 
PBH formation from an overdense region \cite{Shibata:1999zs}.

Assuming a spherically symmetric overdense region,
two conditions must be satisfied for the formation of PBHs from the overdense region.
First, the overdense region must have larger density than a critical value.
Under the linear approximation, 
this is given by curvature perturbation at the horizon crossing 
${\mathcal{R}}(t,r)$ as
\begin{eqnarray}
	\label{eq:condition-C1}
	{\mathcal{R}}(t,0) \gtrsim 0.67~:~~{\mathrm{C1}}.
\end{eqnarray}
The other independent condition is given by
\begin{eqnarray}
	\label{eq:condition-C2}
	-0.72 \lesssim X(r) \lesssim -0.28~:~~{\mathrm{C2}},
\end{eqnarray}
where $X(r)=\frac{r}{2}\frac{d{\mathcal{R}}}{dr}$ at the horizon crossing.
The latter condition requires that the excess mass around the overdense regin
is sufficiently large.
The initial mass fraction $\beta$ of PBHs can be identified with the probability
$P({\mathrm{C1}}\cap{\mathrm{C2}})$.
Assuming that the density perturbation has a Gaussian probability distribution,
the probability $P({\mathrm{C1}})$ is given by
\begin{eqnarray}
	P({\mathrm{C1}}) = \int_{0.67}
	\frac{1}{\sqrt{2\pi}\sigma_{\mathcal{R}}}
	\exp\left( -\frac{\delta^2}{2\sigma^2_{\mathcal{R}}} \right) d\delta,
\end{eqnarray}
where $\sigma_{\mathcal{R}}$ is the variance of ${\mathcal{R}}$ at the horizon crossing.
For the spectrum shown in Fig.\ref{fig:spectrum}-(a), 
we can get $\sigma_{\mathcal{R}}=1.9$, which results in $\beta \simeq 0.35$.

On the other hand, under the presence of a peak with ${\mathcal{R}}(t,0) = 0.67$,
the conditional probability of $({\mathrm{C2}})$ is found to be 0.5.
Consequently, the initial mass fraction of PBHs is estimated to be
\begin{eqnarray}
	\beta \simeq P({\mathrm{C1}}\cap{\mathrm{C2}}) \sim 0.18.
\end{eqnarray}
This is far larger than the constraint $\beta < 4.6\times10^{-6}$.

If we introduce larger decay rates of inflaton fields $\sigma$ and $\psi$, 
resultant PBH abundance can be acceptably small.
The spectrum shown in Fig.\ref{fig:spectrum}-(b) gives 
$\beta \sim 5.6\times10^{-9}$, which is below the bound $\beta < 4.6\times10^{-6}$.

%%%%%%%%%%%%%%%%%%%%%%%%%%%%%%%%%%%%%%%
\section{Conclusion and Discussion}
%%%%%%%%%%%%%%%%%%%%%%%%%%%%%%%%%%%%%%%

We have reexamined the smooth hybrid new inflation model, which was 
designed to reproduce the running spectral index suggested by 
the WMAP three year result,
by numerical calculation of the perturbation. 
We have confirmed that this model can reproduce 
the spectrum within 1$\sigma$ range of the WMAP three year result.

In addition, we find strong peaks on smaller scales, which 
originate from the amplification of perturbation 
$\delta\phi$ and $\delta\psi$ by parametric resonance.
Since there are interactions between $\sigma,\psi$ and $\phi$, 
this amplified perturbation is transfered to $\delta\phi$.
$\delta\phi$ begins large-amplitude and long-period oscillation, 
and freezes out at the horizon crossing, determining the primordial 
curvature perturbation.
The resultant curvature perturbation depends on the phase of 
oscillating $\delta\phi$, so that the resonant peak consists 
of several steep peaks and valleys.
The strength of the resultant peak can be controlled by the decay 
rate of the first inflaton, and the comoving scale of the peak can be 
controlled by the scale factor at the beginning of oscillatory phase.

Such a peak in the spectrum of primordial curvature perturbation 
can be a source of PBHs.
Due to the strong peak, mass of generated PBHs can be estimated 
by the horizon mass at the horizon crossing of the scale 
which corresponds to the resonant peak.

\acknowledgments{ This work is partially supported by the JSPS
Grant-in-Aid for Scientific Research No.\ 18740157 (M.Y.) and No.\
16340076 (J.Y.). M.Y.\ is supported in part by the project of the
Research Institute of Aoyama Gakuin University.}


\begin{thebibliography}{10}
%%%%%%  Introduction  %%%%%%%%
%\cite{Spergel:2003cb}
\bibitem{Spergel:2003cb}
D.~N.~Spergel {\it et al.}  [WMAP Collaboration],
%``First Year Wilkinson Microwave Anisotropy Probe (WMAP) Observations:
%Determination of Cosmological Parameters,''
Astrophys.\ J.\ Suppl.\  {\bf 148}, 175 (2003).
%[arXiv:astro-ph/0302209].
%%CITATION = ASTRO-PH 0302209;%%

%\cite{Peiris:2003ff}
\bibitem{Peiris:2003ff}
H.~V.~Peiris {\it et al.},
%``First year Wilkinson Microwave Anisotropy Probe (WMAP) observations:
%Implications for inflation,''
Astrophys.\ J.\ Suppl.\  {\bf 148}, 213 (2003).
%[arXiv:astro-ph/0302225].
%%CITATION = ASTRO-PH 0302225;%%

%%\cite{Slosar:2004xj}
\bibitem{Slosar:2004xj}
A.~Slosar and U.~Seljak,
% ``Assessing the effects of foregrounds and sky removal in WMAP",
Phys.\ Rev.\ D {\bf{70}}, 083002 (2004).
%[arXiv:astro-ph/0404567].
%%CITATION = ASTRO-PH 0404567;%%"

%\cite{Spergel:2006hy}
\bibitem{Spergel:2006hy}
D.~N.~Spergel {\it et al.}, [arXiv:astro-ph/0603449].
%``Wilkinson Microwave Anisotropy Probe (WMAP) three year results: Implications
%for cosmology,''
%arXiv:astro-ph/0603449.
%%CITATION = ASTRO-PH 0603449;%%

%\cite{Seljak:2004xh}
\bibitem{Seljak:2004xh}
U.~Seljak {\it et al.}  [SDSS Collaboration],
% ``Cosmological parameter analysis including SDSS Ly-alpha forest and  galaxy
%bias: Constraints on the primordial spectrum of fluctuations,  neutrino mass,
%and dark energy,''
Phys.\ Rev.\ D {\bf 71}, 103515 (2005).
%[arXiv:astro-ph/0407372].
%%CITATION = ASTRO-PH 0407372;%%

%%\cite{Ballesteros:2005eg}
\bibitem{Ballesteros:2005eg}
G.~Ballesteros, J.~A.~Casas and J.~R.~Espinosa,
%%``Running spectral index as a probe of physics at high scales",
JCAP. {\bf 0603}, 001 (2006).
%arXiv:[hep-ph/0601134].
%%CITATION = HEP-PH 0601134;%%

%%\cite{Chen:2004nx}
\bibitem{Chen:2004nx}
C.-Y.~Chen, B.~Feng, X.-L.~Wang, and Z.-Y.~Yang,
%%``Reconstructing large running-index inflaton potentials",
Class.\ Quant.\ Grav.\ {\bf 21}, 3223 (2004).
%%arXiv:[astro-ph/0404419].
%%CITATION = ASTRO-PH 0404419;%%

%\cite{Kawasaki:2003zv}
\bibitem{Kawasaki:2003zv}
M.~Kawasaki, M.~Yamaguchi, and J.~Yokoyama,
%``Inflation with a running spectral index in supergravity,''
Phys.\ Rev.\ D {\bf 68}, 023508 (2003).
%[arXiv:hep-ph/0304161].
%%CITATION = HEP-PH 0304161;%%

%\cite{Yamaguchi:2003fp}
\bibitem{Yamaguchi:2003fp}
M.~Yamaguchi and J.~Yokoyama,
%``Chaotic hybrid new inflation in supergravity with a running spectral
%index,''
Phys.\ Rev.\ D {\bf 68}, 123520 (2003).
%[arXiv:hep-ph/0307373].
%%CITATION = HEP-PH 0307373;%%

%\cite{Yamaguchi:2004tn}
\bibitem{Yamaguchi:2004tn}
M.~Yamaguchi and J.~Yokoyama,
%``Smooth hybrid inflation in supergravity with a running spectral index  and
%early star formation,''
Phys.\ Rev.\ D {\bf 70}, 023513 (2004).
%[arXiv:hep-ph/0402282].
%%CITATION = HEP-PH 0402282;%%

%\cite{Lazarides:1995vr}
\bibitem{Lazarides:1995vr}
G.~Lazarides and C.~Panagiotakopoulos,
%``Smooth hybrid inflation,''
Phys.\ Rev.\ D {\bf 52}, R559 (1995).
%%CITATION = HEP-PH 9506325;%%

%\cite{Kumekawa:1994gx}
\bibitem{Kumekawa:1994gx}
  K.~Kumekawa, T.~Moroi, and T.~Yanagida,
  %``Flat potential for inflaton with a discrete R invariance in supergravity,''
  Prog.\ Theor.\ Phys.\  {\bf 92}, 437 (1994).
%  [arXiv:hep-ph/9405337].
  %%CITATION = HEP-PH 9405337;%%
  %\cite{Izawa:1996dv}

\bibitem{Izawa:1996dv}
  K.~I.~Izawa and T.~Yanagida,
  %``Natural new inflation in broken supergravity,''
  Phys.\ Lett.\ B {\bf 393}, 331 (1997).
%  [arXiv:hep-ph/9608359].
  %%CITATION = HEP-PH 9608359;%%
%\cite{Ibe:2006fs}

\bibitem{Ibe:2006fs}
  M.~Ibe, K.~I.~Izawa, Y.~Shinbara, and T.~T.~Yanagida,
  %``Minimal supergravity, inflation, and all that,''
  [arXiv:hep-ph/0602192].
  %%CITATION = HEP-PH 0602192;%%
 %\cite{Asaka:1999jb}

%\cite{Kofman:1994rk}
\bibitem{Kofman:1994rk}
L.~Kofman, A.~D.~Linde, and A.~A.~Starobinsky,
%``Reheating after inflation,''
Phys.\ Rev.\ Lett.\  {\bf 73}, 3195 (1994).
%%CITATION = HEP-TH 9405187;%%

%\cite{Shtanov:1994ce}
\bibitem{Shtanov:1994ce}
Y.~Shtanov, J.~H.~Traschen, and R.~H.~Brandenberger,
%``Universe reheating after inflation,''
Phys.\ Rev.\ D {\bf 51}, 5438 (1995).
%%CITATION = HEP-PH 9407247;%%

%\cite{Kofman:1997yn}
\bibitem{Kofman:1997yn}
L.~Kofman, A.~D.~Linde, and A.~A.~Starobinsky,
%``Towards the theory of reheating after inflation,''
Phys.\ Rev.\ D {\bf 56}, 3258 (1997).
%%CITATION = HEP-PH 9704452;%%

%\cite{Yamaguchi:2005}
\bibitem{Yamaguchi:2005}
M. Yamaguchi and J. Yokoyama,
[arXiv:hep-ph/0512318].

%\cite{Kawasaki:1998vx}
\bibitem{Kawasaki:1998vx}
  M.~Kawasaki and T.~Yanagida,
  %``Primordial black hole formation in supergravity,''
  Phys.\ Rev.\ D {\bf 59}, 043512 (1999).
%  [arXiv:hep-ph/9807544].
  %%CITATION = HEP-PH 9807544;%%

%\cite{Salopek:1989qh}
\bibitem{Salopek:1989qh}
  D.~S.~Salopek, J.~R.~Bond, and J.~M.~Bardeen,
  %``Designing density fluctuation spectra in inflation,''
  Phys.\ Rev.\ D {\bf 40}, 1753 (1989).

\bibitem{MathieuFunctions}
   N.W.MacLachlan, {\em{Theory and Application of Mathieu Functions}}, 
   Dover, New York (1961).
  
%\cite{Felder:2000hj}
\bibitem{Felder:2000hj}
  G.~N.~Felder, J.~Garcia-Bellido, P.~B.~Greene, L.~Kofman, A.~D.~Linde, and I.~Tkachev,
  %``Dynamics of symmetry breaking and tachyonic preheating,''
  Phys.\ Rev.\ Lett.\  {\bf 87}, 011601 (2001).
  %[arXiv:hep-ph/0012142].
  %%CITATION = HEP-PH 0012142;%% 

%\cite{Copeland:2002ku}
\bibitem{Copeland:2002ku}
  E.~J.~Copeland, S.~Pascoli, and A.~Rajantie,
  %``Dynamics of tachyonic preheating after hybrid inflation,''
  Phys.\ Rev.\ D {\bf 65}, 103517 (2002).
  %[arXiv:hep-ph/0202031].
  %%CITATION = HEP-PH 0202031;%%

%\cite{Green:1997sz}
\bibitem{Green:1997sz}
  A.~M.~Green and A.~R.~Liddle,
  %``Constraints on the density perturbation spectrum from primordial black  holes,''
  Phys.\ Rev.\ D {\bf 56}, 6166 (1997).

%\cite{Yokoyama:1999xi}
\bibitem{Yokoyama:1999xi}
  J.~Yokoyama,
  %``Formation of primordial black holes in inflationary cosmology,''
  Prog.\ Theor.\ Phys.\ Suppl.\ {\bf 136}, 338 (1999).

%\cite{Shibata:1999zs}
\bibitem{Shibata:1999zs}
  M.~Shibata and M.~Sasaki, 
  %``Formation of primordial black holes in inflationary cosmology,''
  Phys.\ Rev.\ D{\bf 60}, 084002 (1999).

\end{thebibliography}
\end{document}